\documentclass[prl,twocolumn,showpacs,showkeys]{revtex4-1}
\usepackage{graphicx}
\usepackage{dcolumn}
\usepackage{longtable}
\usepackage{amsmath}
\usepackage{amssymb}
\usepackage{natbib}
\usepackage{array}
\usepackage{float}

\begin{document}
	\title{Non-relativistic conformal symmetry in nuclear structure and relations with the unitarity limit}
	\date{\today}
	\author{P E Georgoudis}
	\affiliation{Grand Accelerateur National d'Ions Lourds, CEA/IRFU, BP 55027, F-14076 Caen Cedex 5, France}
\email{panagiotis.georgoudis@ganil.fr}
\begin{abstract}
The Interacting Boson Model of nuclear structure is introduced to the unitarity limit. Non relativistic conformal symmetry creates a scale invariant state from which the stationary states of atomic nuclei are obtained. A brief discussion of the nature of the conformal field theory is included along with the BCS-BEC crossover and hydrodynamic properties.
\end{abstract}

	\maketitle
The current theoretical description of nuclear collective states in atomic nuclei is not accompanied by an explicit manifestation of the strong interactions at the level of mesons or of Quantum Chromo Dynamics.  However, atomic nuclei exhibit symmetries in their collective states that are hosted in the group theoretical framework of the Interacting Boson Model \cite{IBM} with remarkable phenomenological success. One may then raise the question about the relation of the symmetries of the IBM with the symmetries of the strong interactions, either with the SU(N) gauge group or with conformal symmetry as the classical limit of QCD. This problem reflects in part the more general one of the understanding of the relation between the symmetries manifested in stationary states of subatomic structures, of which atomic nuclei are an example, with the symmetries of the fundamental interactions as they are manifested in scattering states between the constituents of those structures.  
	
In nuclear structure, the occurrence of conformal invariance is implied through the E(5) symmetry \cite{E(5)} that emerges at the critical point of a 2nd order Quantum Phase Transition between the U(5) and O(6) dynamical symmetry limits of the IBM. Three different subgroup chains of the U(6) group consist of the dynamical symmetry limits of the IBM and correspond to shapes of the nuclear surface. These are the U(5) limit (spherical), the O(6) limit  ($\gamma$ unstable) and the SU(3) limit (axially symmetric). U(6) is the symmetry of a six dimensional harmonic oscillator. A spin zero $s$ boson and a spin two $d$ boson are its building blocks. An oscillator length is obtainable first by the identification of the $s$ and $d$ bosons with valence nucleon pairs of total angular momentum zero and two respectively and second by the phenomenological application of a subgroup chain to a total number of bosons (valence nucleon pairs) which determine an atomic nucleus. It is of interest to obtain an oscillator length for the IBM via a procedure which involves a scattering length that characterizes the scattering of the interacting constituents which participate in a nuclear collective state. 
	
In non-relativistic quantum mechanics fermions at unitarity \cite{Randeria, Son} manifest non relativistic conformal symmetry and support a quantum critical point. This is the unitarity limit - an otherwise scattering problem at the limit of infinite scattering length, which has been applied in light nuclei in the framework of the No Core Shell Model \cite{Barrett}. The unitarity limit supports a bound state of zero energy and since it reflects universality \cite{Konig} one may expect that its application in nuclear structure should not distinguish between light and heavy nuclei.

 This paper introduces the connection of nuclear collective stationary states that are labelled by the O(6) quantum number of the IBM with the unitarity limit. The first outcome is the emergence of the SO(2,1) conformal group in the IBM. This symmetry permits to define an equation for the scaling behavior of the IBM wavefunctions at unitarity. The second is the introduction of the relation of the IBM at unitarity with a non-relativistic conformal field theory, with the BCS-BEC crossover and finally with hydrodynamics.

Take a system of N fermions with 3N coordinates $r_{i}$ subjected to the free Hamiltonian
\begin{equation}\label{e1}
H=\sum_{i}\frac{{p}^{2}_{i}}{2m},
\end{equation}	
with an overall wavefunction $\psi({r}_{1},...{r}_{N})$ satisfying boundary conditions
\begin{equation}\label{e2}
\lim_{{r}_{i}\rightarrow {r}_{j}} \psi({r}_{1},...{r}_{N})= \frac{C}{|{r}_{i}-{r}_{j}|}+ O(|{r}_{i}-{r}_{j}|).
\end{equation}
These are fermions at unitarity. The physics in this case is said to be universal in the sense that the scattering length between the particles goes to infinity leaving no characteristic length scale in the boundary condition. This limit resembles the case of the nucleon-nucleon interaction with the short-range repulsion. 

An interesting relation which reflects the unitarity limit is the zero-energy solution of the Schrodinger equation
\begin{equation}\label{e3}
\sum_{i} \frac{\partial^{2}}{\partial r^{2}_{i}}\psi({r}_{1},...{r}_{N})=0,
\end{equation}
with the scaling behavior
\begin{equation}\label{e4}
\psi({r}_{1},...{r}_{N})=R^{\nu} \psi(\Omega_{k}).
\end{equation} 
It is precisely this scaling behavior that is of interest in relation to nuclear collective effects. Here, $\nu$ is a scaling exponent and $R$ has dimensions of length setting an overall scale for the distances between the fermions within a trapping potential. These are standard hyperspherical coordinates with $R^{2}=\sum_{i}r^{2}_{i}$ and the angles $\Omega_{k}$ reflect the ratios between the $r_{i}$.

I would like now to discuss the similarity between the Schrodinger Equation of the $U(6) \supset O(6)$ limit of the IBM \cite{Mexico}  and that of the non-relativistic Schrodinger equation for N particles in hyperspherical coordinates \cite{WC, BEC} confided in a harmonic oscillator trap. These two equations are

\begin{widetext}
\begin{equation}\label{Mex}
\left[ \frac{1}{2} \left( -\frac{1}{R^{5}} \frac{\partial}{\partial R}R^{5} \frac{\partial}{\partial R} + \frac{ \sigma ( \sigma+ 4 ) }{R^{2}}+R^{2} \right) \right] F^{\sigma}_{J}(R)= 
\left(N_{b} + \frac{6}{2} \right)F^{\sigma}_{J}(R),
\end{equation}
\begin{equation}\label{WC}
-\frac{\hbar^{2}}{2m} \left(  \frac{1}{R^{3N-1}} \frac{\partial}{\partial R}R^{3N-1} \frac{\partial}{\partial R} \right)\Phi(R) + 
\left( \frac{\hbar^{2} \Lambda}{2mR^{2}} +\frac{1}{2} m\omega^{2} R^{2} \right)\Phi(R) =E \Phi(R),
\end{equation}
\end{widetext}
respectively. In Eq (\ref{WC}) we didn't take into account the center of mass like in \cite{BEC} and its angular wavefunctions obey the relation
\begin{equation}\label{e7}
\Lambda^{2} Y_{\lambda \mu}(\Omega)=\lambda(\lambda+3N-2)Y_{\lambda \mu}(\Omega) \equiv \Lambda Y_{\lambda \mu}(\Omega).
\end{equation}
For $N=2$ the angular eigenvalues are those of the Casimir of the O(6) and conforming with the IBM notation they are written as $\sigma(\sigma+4)$. The unitarity limit is introduced in Eq (\ref{WC}) by setting $\omega=0$ and $E=0$ \cite{WC}. The state at unitarity is scale invariant. Then, at any $N$ the trapping of the scattered fermions at unitarity preserves scale invariance and furthermore manifests the SO(2,1) conformal algebra \cite{Pitaevskii}
\begin{equation}\label{e8}
[H,D]=-2iH, \quad [K,D]=2iK, \quad [K,H]=i \omega^{2} D,
\end{equation}
with $H$ the free Hamiltonian of Eq (\ref{e1}), $K=(1/2) m \omega^{2} R^{2}$ is the special conformal generator and the dilatation operator is $D=R\partial_{R}$ with eigenvalue $\nu$. At unitarity, the zero energy solution (\ref{e3}) is translated to the eigenvalues of the angular operator \cite{WC}
\begin{equation}\label{e9}
\Lambda=\nu (\nu+3N-2).
\end{equation}
 For $N=2$, $\sigma=\nu$ i.e the O(6) is preserved at unitarity. Now, Eq (\ref{Mex}) represents the $U(6) \supset O(6)$ limit of the IBM with wavefunctions
\begin{equation}\label{rwfs}
F^{\sigma}_{J}(R)=R^{\sigma} L^{\sigma+2}_{J}(R^{2}) e^{-R^{2}/2},
\end{equation}
with $ L^{\sigma+2}_{J}(R)$ the associated Laguerre polynomial and $J=(N_{b}-\sigma)/2$ is constrained by the total number of bosons $N_{b}$ in the $U(6)$ symmetry. The total wavefunction reads
\begin{equation}\label{e12}
\Psi(R,\chi,\Omega_{4})= F^{\sigma}_{J}(R) g^{\sigma}_{\tau}(\chi) Y^{\tau}_{L}(\Omega_{4}).
\end{equation}
 The radial wavefunction $F^{\sigma}_{J}(R)$ describes excitations with respect to the total number of bosons. In the geometric scheme for the representation of the six dimensional harmonic oscillator of the IBM \cite{Mexico, Panos}, $R=\sqrt{\beta^{2}+q_{0}^{2}}$, with $\beta$ the quadrupole deformation of the nuclear surface and $q_{0}$ the $s$ boson coordinate that is transverse to the five dimensional quadrupole plane. $R$ is taken as the radius of the nuclear surface including its quadrupole deformation. This is a matching condition with the expression of $R$ as the radius of the trap in the hyperspherical coordinates. For a specific $N_{b}$, the wavefunctions  $g^{\sigma}_{\tau}(\chi)$ are related with the $\beta$ vibrations in the O(6) limit and $Y^{\tau}_{L}(\Omega_{4})$ are the spherical harmonics of the O(5) symmetry. A characteristic oscillator length $a_{ho}$ is fixed for the IBM by introducing a characteristic mass $m$ and frequency $\omega$ in Eq (\ref{Mex}). The radial wavefunctions then read
\begin{equation}\label{aho}
F^{\sigma}_{J}(R)=R^{\sigma} L^{\sigma}_{J}(R^{2}/a^{2}_{ho}) e^{-R^{2}/2a^{2}_{ho}},
\end{equation}
and the spectrum is
\begin{equation}\label{spc}
E_{\sigma,J}=\left(\sigma+2 J +\frac{6}{2} \right) \hbar \omega.
\end{equation}
Equations (\ref{Mex}) and (\ref{WC}) coincide for $N=2$ and making the correspondence $E=(N_{b}+6/2)\hbar \omega$. Upon switching off ($\omega=0$) the trapping potential, unitarity is achieved at $E=0$. This is a limit in which the lower bound of $(6/2)\hbar \omega$ from the $U(6)$ is absent. In this case the $O(6)$ symmetry is preserved with the $\sigma$ label to play the role of the eigenvalue of the dilatation operator $R\partial_{R}$. 
In the hyperspherical Eq (\ref{WC}) the number of particles $N$ is finally introduced as a gauge factor in the radial wavefunction which can be chosen up to our freedom to eliminate the first derivative. Setting $\Phi(R)=R^{(1-3N)/2}F(R)$, Eq (\ref{WC}) can be cast in the form
\begin{equation}\label{WC2}
\left(-\frac{\hbar^{2}}{2m}  \frac{\partial^{2}}{\partial R^{2}} +  \frac{\hbar^{2} (p(p+4))}{2mR^{2}} +\frac{1}{2} m\omega^{2} R^{2} \right) F(R) =E F(R),
\end{equation}
in which the parameter $p$ is defined by 
\begin{equation}
p(p+4)=\sigma(\sigma+4)+\frac{(3N-1)(3N-3)}{4}.
\end{equation}
The solutions are the same as for $N=2$ but with $p$ in place of $\sigma$ in the wavefunctions Eq (\ref{aho}) and in the spectrum Eq (\ref{spc}). The O(6) symmetry is present in both cases like in the relevant solutions of the O(5) Schrodinger equation \cite{Elliott}. For $N=2$, the eigenvalue of the dilatation operator is $\nu=\sigma$ while for $N>2$ is
\begin{equation}
\nu=\frac{1-3N}{2}+p.
\end{equation}
On the one hand, the hyperspherical problem for $N\geq 2$ is reduced to the two body problem through the $O(6)$ symmetry. On the other, the unitarity limit of the IBM is now straightforward for any N which serves for the interpretation of stationary collective states of nuclear structure as trapped states of a scattering problem. The character of fermions at unitarity is in principle indeterminate. Let's interpret a free space eigenstate $\psi^{0}_{\nu}$ as a set of N nucleons out of structure. For $N=2$ this dimer of nucleons should not be confused with one valence nucleon pair, but rather for now leave the character of the bosons indeterminate too. We can perform a mapping to states within structure. An eigenstate of an $O(6)$ nucleus is written as
\begin{equation}
|\psi(\sigma,J) \rangle = (L_{+})^{J}e^{-R^{2}/2 a^{2}_{ho}} |\psi^{0}_{\sigma} \rangle, 
\end{equation}
with the spectrum of Eq (\ref{spc}). For $N>2$ the same mapping holds with $p$ in place of $\sigma$. $L_{\pm}$ are defined as
\begin{equation}
\begin{aligned}
L_{+} & =\frac{6}{2}+ D+ \frac{H}{\hbar \omega} - 2K, \\
L_{-} & =-\frac{6}{2}- D+ \frac{H}{\hbar \omega} - 2K.
\end{aligned}
\end{equation}
In terms of $s$ and $d$ bosons, the square of the radius $R^{2}=\beta^{2}+q^{2}_{0}$ gives the special conformal generator 
\begin{equation}
	K=\frac{1}{2}m \omega^{2}\left( (d^{\dagger}+d)(d^{\dagger}+d)+(s^{\dagger}+s)(s^{\dagger}+s) \right). 
\end{equation}
The kinetic term of the O(6) IBM Hamiltonian is denoted here as $H$. 
The SO(2,1) algebra is revealed in the IBM by expressing these generators in their cartesian form. With $q_{i}$ the coordinates of the six dimensional harmonic oscillator \cite{Panos} take as
\begin{equation}
	H=\sum_{i=0}^{5}-\frac{\hbar^{2}}{2m}\frac{\partial^{2}}{\partial q^{2}_{i}}, \quad K=\sum_{i=0}^{5}\frac{1}{2} m \omega^{2} q^{2}_{i}.
\end{equation}
The non-relativistic Schrodinger equation for the $U(6)$ IBM can be built with the operator $H+K$ and Eq (\ref{Mex}) consists of its O(6) limit. Now,
\begin{equation}
	\begin{split}
		& D=\sum_{i}\frac{1}{2}(\partial_{i} q_{i}+q_{i} \partial_{i})=\frac{1}{2}((s-s^{\dagger})(s+s^{\dagger})+ \\
		& (s+s^{\dagger})(s-s^{\dagger})+ (d-d^{\dagger})(d+d^{\dagger})+(d^{\dagger}+d)(d-d^{\dagger})), 
	\end{split}
\end{equation}
and
\begin{equation}
H= -\frac{\hbar^{2}}{2m}((s-s^{\dagger})(s-s^{\dagger})+(d-d^{\dagger})(d-d^{\dagger})).
\end{equation}
The state $|\psi^{0}_{\sigma} \rangle$ is of zero energy and scale invariant. This means that under a scaling transformation of the total distance of the two nucleons from the center of the trap of the form $R \rightarrow R/a_{ho}$, the state scales as
\begin{equation}
\psi^{0}_{\sigma}(R/a_{ho})=\psi^{0}_{\sigma}(R)/a_{ho}^{\sigma},
\end{equation}
with $\sigma$ to play the role of the scaling exponent. Taking a state from the trap by applying the equation
\begin{equation}
L_{-}|\psi \rangle = 0,
\end{equation}
one can obtain the free space eigenstate from the trapped state \cite{WC}. The solution is
\begin{equation}
\psi^{0}_{\sigma}(R)=R^{\sigma} g^{\sigma}_{\tau}(\chi) Y^{\tau}_{L}(\Omega_{4}).
\end{equation}
Applying this state in the O(6) Schrodinger equation (\ref{Mex}) for $\omega=0$ gives indeed $E=0$ by the direct reproduction of $\sigma(\sigma+4)=\sigma(\sigma+4)$ for $J=0$ ($N_{b}=\sigma+2 J$).

The solutions of the IBM with radial wavefunction $F^{\sigma}_{J}(R)=R^{\sigma}$ and energy absolute zero correspond to $N=2$ fermions at unitarity. The solutions with $F^{\sigma}_{J}(R)=R^{p}$ correspond to $N>2$ fermions at unitarity. This means that the radial wavefunctions of Eq (\ref{aho}) at unitarity are
\begin{equation}
F^{\sigma}_{J}(R)=R^{\sigma} L^{\sigma}_{J}(R^{2}/a^{2}_{ho}) e^{-R^{2}/2a^{2}_{ho}} \rightarrow R^{\sigma}.
\end{equation}
Therefore, the scaling behavior of the IBM wavefunctions at unitarity is conveyed to the equation
\begin{equation}\label{1}
\psi({r}_{1},...{r_{N}})=R^{\nu} g^{\sigma}_{\tau}(\chi) Y^{\tau}_{L}(\Omega_{4}).
\end{equation}
If $N=2$, $\nu=\sigma$ while if $N>2$, $\nu=(1-3N)/2+p$. In both cases the O(6) label $\sigma$ is present. These steps formulate the unitarity limit of the IBM. It deserves to be mentioned that relations with the unitarity limit are also achievable when the center of mass is considered. In these cases the exponential of the radial term in Eq (\ref{WC}) is $3N-4$ and the O(6) symmetry emerges for $N=3$. Jacobi coordinates can be used in this case and involve $N-1$ radii. The scaling behavior for $N>3$ follows the same behavior with the correspondent index $p$ instead of $\sigma$.

The interpretation of the unitarity limit of the IBM is better viewed through the picture of defining a scattering problem out of structure and studying its limit at zero energy. Scale invariance is interpreted in the IBM through the O(6) invariance. The first consequence to be discussed is the relation of stationary states of the IBM with scattering. This is not a new proposal for the IBM \cite{Alhassid} but the new element is the unitarity limit. In nuclear scattering the usual formalism involves the concept of the scattering channels with the external region and the internal one in which the compound nucleus is formed \cite{Feshbach}. The boundary between these two regions defines the nuclear radius. We set this boundary to be the nuclear surface. The hyperspherical formalism applies to the external region while the geometric scheme of the IBM applies to the boundary. The unitarity limit of the IBM is represented by Eq (\ref{1}) which combines the two regions. 

The second consequence is related to the underlying non relativistic conformal field theory at unitarity \cite{Son}. Such a theory permits us to define a global approach to nuclear structure. The solutions of the hyperspherical equation (\ref{WC}) are suitable as a basis of the No Core Shell Model for light nuclei. Their distinction from the collective states which are present in heavier nuclei depends on the scaling dimension of the representation of the conformal algebra.  A fermion creation operator in three dimensions with scaling dimension \cite{Son} $\Delta=3/2$ is to be regarded as a nucleon in this application. A No Core Shell Model basis is generated for this creation operator as a three dimensional harmonic oscillator after the mapping from unitarity. The dimer creation operator, the coupling of two nucleons, has scaling dimension $\Delta=2$ and after the mapping from unitarity generates a collective basis which spans the representations of the $U(6) \supset O(6)$ limit of the IBM. These creation operators act on the vacuum to create the scale invariant state $\psi^{0}_{\nu}$. The stationary states are obtained after the mappings with the $L_{\pm}$ operators.  In both cases the fundamental frequency $\omega$ is the same while in the latter case the $L_{\pm}$ operators create a ladder of states separated by $2\hbar \omega$. 

Upon attributing to the fermions at unitarity an $SU(2)$ gauge group like in \cite{Son}, the theory is formally similar with that of Mukerjee and Nambu \cite{Nambu} for the IBM in which the transverse excitations of the gauge group refer to the $\sigma$ meson while the longitudinal ones refer to  the $\pi$ meson. Both theories evaluate bubble diagrams between the fermions, in infinite nuclear matter these diagrams generate collectivity, and the order of approximation provides the cutoff of the renormalization group. Now, the special conformal generator $K$ traps the nucleons at unitarity in a scale determined by the nuclear radius. This aspect of conformality may provide a symmetry solution to the problem of the finiteness of the Hilbert space of atomic nuclei in the field theory of Makerjee and Nambu for the IBM. 

The third consequence is the introduction of the BCS-BEC crossover in nuclear structure and in the IBM as a feature of the unitarity limit. The usual BCS interpretation of the valence nucleon pairs does not seem to be relevant in this case since unitarity represents a scale invariant state out of structure. Rather, this case resembles the two channel models in molecules with the bosonic state to be formed in the closed channel (internal region) and the fermionic in the open channel (external region). In relation with the underlying field theory at unitarity which manifests the crossover, the introduction of a gauge group acts as the link with the strong interactions. A Bose Einstein Condensation limit of the IBM in such a case bears a resemblance with a large N limit of an SU(N) gauge group like in a Skyrme model. Its crossover with a BCS theory, an example of such a BCS for the SU(2) gauge group has been elaborated in \cite{Nambu} but lacks conformal symmetry, may bring the IBM to the regime of color superconductivity.

The fourth and final consequence is related to hydrodynamics. The unitarity limit of the $U(6) \supset O(6)$ limit of the IBM with $\omega=0$ and $E=0$ is geometrically equivalent with the limit of infinite Radius in the O(6) limit \cite{Panos}, in both cases the curvature is absent. Such a limit serves for the the relation with the E(5) symmetry and the collective model which lives at the limit of infinite radius. The unitarity limit hosts fluid properties \cite{Randeria} and serves for the relation of the IBM with hydrodynamics. The explicit comparison of the nuclear collective modes of motion with the oscillations of an irrotational and non viscous fluid may be succeeded by the unitarity limit of the IBM.

\begin{acknowledgments}
This work is supported by the European Union's H2020 program, Marie Sklodowska Curie Actions, Grant Agreement No 793900-GENESE17.
\end{acknowledgments}

\end{document}